\documentclass[twocolumn,floatfix,showpacs]{revtex4}
\usepackage{graphicx}
\usepackage{amsmath}
\usepackage{bm}
\begin{document}

\title{Detection of valley polarization in graphene by a superconducting contact}
\author{A. R. Akhmerov}
\affiliation{Instituut-Lorentz, Universiteit Leiden, P.O. Box 9506, 2300 RA Leiden, The Netherlands}
\author{C. W. J. Beenakker}
\affiliation{Instituut-Lorentz, Universiteit Leiden, P.O. Box 9506, 2300 RA Leiden, The Netherlands}
\date{December 2006}
\begin{abstract}
Because the valleys in the band structure of graphene are related by time-reversal symmetry, electrons from one valley are reflected as holes from the other valley at the junction with a superconductor. We show how this Andreev reflection can be used to detect the valley polarization of edge states produced by a magnetic field. In the absence of intervalley relaxation, the conductance $G_{\rm NS}=(2e^{2}/h)(1-\cos\Theta)$ of the junction on the lowest quantum Hall plateau is entirely determined by the angle $\Theta$ between the valley isospins of the edge states approaching and leaving the superconductor. If the superconductor covers a single edge, $\Theta=0$ and no current can enter the superconductor. A measurement of $G_{\rm NS}$ then determines the intervalley relaxation time. 
\end{abstract}
\pacs{74.45.+c, 73.23.-b, 73.43.-f, 73.50.Jt}
\maketitle

The quantized Hall conductance in graphene exhibits the half-integer quantization $G_{\rm H}=(n+\frac{1}{2})(4e^{2}/h)$ characteristic of massless Dirac fermions \cite{Nov05,Zha05}. The lowest plateau at $2e^{2}/h$ extends to zero carrier density because there is no gap between conduction and valence bands, and it has only a twofold spin degeneracy because it lacks the valley degeneracy of the higher plateaus. The valley degeneracy of the lowest Landau level is removed at the edge of the carbon monolayer, where the current-carrying states at the Fermi level are located. Depending on the crystallographic orientation of the edge, the edge states may lie fully within a single valley, or they may be a linear combination of states from both valleys \cite{Bre06,Aba06}. The type of valley polarization remains hidden in the Hall conductance, which is insensitive to edge properties.

Here we propose a method to detect the valley polarization of quantum Hall edge states, using a superconducting contact as a probe. In the past,  experimental \cite{Tak98a,Moo99,Uhl00,Ero05} and theoretical \cite{Tak98b,Hop00,Asa00,Cht01,Gia05} studies of the quantum Hall effect with superconducting contacts have been carried out in the context of semiconductor two-dimensional electron gases. The valley degree of freedom has not appeared in that context. In graphene, the existence of two valleys related by time-reversal symmetry plays a key role in the process of Andreev reflection at the normal-superconducting (NS) interface \cite{Bee06}. A nonzero subgap current through the NS interface requires the conversion of an electron approaching in one valley into a hole leaving in the other valley. This is suppressed if the edge states at the Fermi level lie exclusively in a single valley, creating a sensitivity of the conductance of the NS interface to the valley polarization.

\begin{figure}[tb]
\centerline{\includegraphics[width=0.9\linewidth]{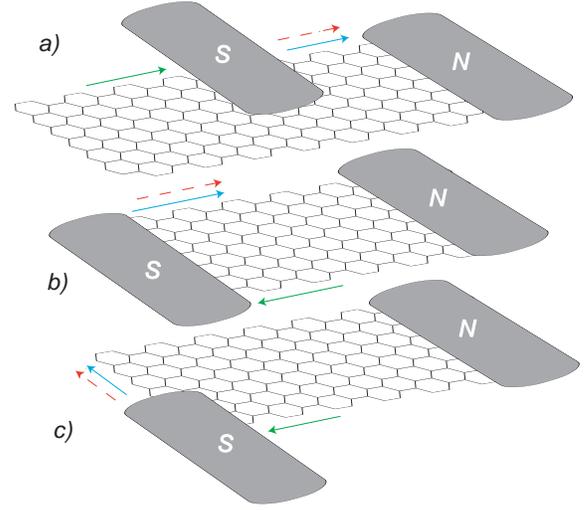}}
\caption{\label{NS_layout}
Three diagrams of a graphene sheet contacted by one normal-metal (N) and one superconducting (S) electrode. Edge states approaching and leaving the superconductor are indicated by arrows. The solid line represents an electron state (green: isospin $\bm{\nu}_{1}$; blue: isospin $\bm{\nu}_{2}$), and the dashed line represents a hole state (red: isospin $-\bm{\nu}_{2}$).
}
\end{figure}

Allowing for a general type of valley polarization, we calculate that the two-terminal conductance $G_{\rm NS}$ (measured between the superconductor and a normal-metal contact) is given by
\begin{equation}
G_{\rm NS}=\frac{2e^{2}}{h}(1-\cos\Theta),\label{GNSresult}
\end{equation}
when the Hall conductance $G_{\rm H}=2e^{2}/h$ is on the lowest plateau \cite{Pra06}. Here $\cos\Theta=\bm{\nu}_{1}\cdot\bm{\nu}_{2}$ is the cosine of the angle between the valley isospins $\bm{\nu}_{1},\bm{\nu}_{2}$ of the states along the two graphene edges connected by the superconductor (see Fig.\ \ref{NS_layout}). If the superconductor covers a single edge (Fig.\ \ref{NS_layout}a), then $\Theta=0\Rightarrow G_{\rm NS}=0$ --- no current can enter into the superconductor without intervalley relaxation. If the superconductor connects different edges (Figs.\ \ref{NS_layout}b,c) then $G_{\rm NS}$ can vary from $0$ to $4e^{2}/h$ --- depending on the relative orientation of the valley isospins along the two edges.

We start our analysis from the Dirac-Bogoliubov-De Gennes (DBdG) equation \cite{Bee06}
\begin{equation}
\begin{pmatrix}
H-\mu&\Delta\\
\Delta^{\ast}&\mu-THT^{-1}
\end{pmatrix}
\Psi
=\varepsilon\Psi, \label{DBdG}
\end{equation}
with $H$ the Dirac Hamiltonian, $\Delta$ the superconducting pair potential, and $T$ the time reversal operator. The excitation energy $\varepsilon$ is measured relative to the Fermi energy $\mu$. Each of the four blocks in Eq.\ (\ref{DBdG}) represents a $4\times 4$ matrix, acting on $2$ sublattice and $2$ valley degrees of freedom. The wave function $\Psi=(\Psi_{e},\Psi_{h})$ contains a pair of $4$-dimensional vectors $\Psi_{e}$ and $\Psi_{h}$ that represent, respectively, electron and hole excitations.

The pair potential $\Delta$ is isotropic in both the sublattice and valley degrees of freedom. It is convenient to choose a ``valley isotropic'' basis such that the Hamiltonian $H$ is isotropic in the valley degree of freedom \cite{note1},
\begin{eqnarray}
H&=&v\begin{pmatrix}
(\bm{p}+e\bm{A})\cdot\bm{\sigma}&0\\
0&(\bm{p}+e\bm{A})\cdot\bm{\sigma}
\end{pmatrix}\nonumber\\
&=&v\tau_{0}\otimes(\bm{p}+e{\bm A})\cdot\bm{\sigma},\label{Hdef}
\end{eqnarray}
with $v$ the Fermi velocity, $\bm{p}=(\hbar/i)(\partial/\partial x,\partial/\partial y)$ the canonical momentum operator in the $x$-$y$ plane of the graphene layer and $\bm{A}$ the vector potential corresponding to a perpendicular magnetic field $B$. The Pauli matrices $\sigma_{i}$ and $\tau_{i}$ act on the sublattice and valley degree of freedom, respectively (with $\sigma_{0}$ and $\tau_{0}$ representing the $2\times 2$ unit matrix). The time reversal operator in the valley isotropic basis reads
\begin{equation}
T=\begin{pmatrix}
0&i\sigma_{y}\\
-i\sigma_{y}&0
\end{pmatrix}{\cal C}=-(\tau_{y}\otimes\sigma_{y}){\cal C},\label{Tdef}
\end{equation}
with ${\cal C}$ the operator of complex conjugation. For later use we note that the particle current operator $\bm{J}=(\bm{J}_{e},\bm{J}_{h})$ has electron and hole components
\begin{equation}
{\bm J}=v(\tau_{0}\otimes{\bm\sigma},-\tau_{0}\otimes{\bm\sigma}).\label{jdef}
\end{equation}

Substitution of Eqs.\ (\ref{Hdef}) and (\ref{Tdef}) into Eq.\ (\ref{DBdG}) gives the DBdG equation in the valley isotropic form
\begin{eqnarray}
&&\begin{pmatrix}
H_{+}-\mu&\Delta\\
\Delta^{\ast}&\mu-H_{-}
\end{pmatrix}
\Psi
=\varepsilon\Psi, \label{DBdG2}\\
&&H_{\pm}=v\tau_{0}\otimes(\bm{p}\pm e{\bm A})\cdot\bm{\sigma}.\label{Hpmdef}
\end{eqnarray}
We seek a solution in the normal region (where $\Delta\equiv 0$), at energies below the excitation gap $\Delta_{0}$ in the superconductor. Electron and hole excitations cannot propagate into the superconductor at subgap energies, and the magnetic field confines them in the normal region to within a magnetic length $l_{m}=\sqrt{\hbar/eB}$ of the edge. We consider separately the edge states along the insulating edge of the graphene layer and along the interface with the superconductor. The edges are assumed to be smooth on the scale of $l_{m}$, so that they may be treated locally as a straight line. We also assume that the superconducting coherence length $\xi_{0}=\hbar v/\Delta_{0}$ is small compared to $l_{m}$, so that the effect of the magnetic field on the superconductor can be neglected.

The edge states at the insulating and superconducting boundaries are different because of the different boundary conditions. Using only the condition of particle current conservation, these have the general form \cite{McC04}
\begin{equation}
\Psi={\cal M}\Psi,\label{calMdef}
\end{equation}
with ${\cal M}$ a unitary and Hermitian matrix that anticommutes with the particle current operator:
\begin{equation}
{\cal M}={\cal M}^{\dagger},\;\;{\cal M}^{2}=1,\;\;{\cal M}(\bm{n}\cdot\bm{J})+(\bm{n}\cdot\bm{J}){\cal M}=0.\label{Mconditions}
\end{equation}
The unit vector $\bm{n}$ lies in the $x$-$y$ plane, perpendicular to the boundary and pointing outward.

At the NS interface the matrix ${\cal M}$ is given by \cite{Tit06}
\begin{equation}
{\cal M}=
\begin{pmatrix}
0&M_{\rm NS}\\
M_{\rm NS}^{\dagger}&0
\end{pmatrix},\;\;
M_{\rm NS}=\tau_{0}\otimes e^{i\phi+i\beta\bm{n}\cdot\bm{\sigma}},\label{MNSdef}
\end{equation}
with $\beta=\arccos(\varepsilon/\Delta_{0})\in(0,\pi)$ determined by the order parameter $\Delta=\Delta_{0}e^{i\phi}$ in the superconductor. 

The insulating (I) edge does not mix electrons and holes, so ${\cal M}$ is block-diagonal with electron block $M_{\rm I}$ and hole block $TM_{\rm I}T^{-1}$. The boundary condition is determined by confinement on the scale of the lattice constant $a\ll l_{m}$, so it should preserve time-reversal symmetry. This implies that $M_{\rm I}$ should commute with $T$. The most general matrix that also satisfies Eq.\ (\ref{Mconditions}) is given by \cite{note2}
\begin{equation}
{\cal M}=
\begin{pmatrix}
M_{\rm I}&0\\
0&M_{\rm I}
\end{pmatrix},\;\;
M_{\rm I}=(\bm{\nu}\cdot\bm{\tau})\otimes(\bm{n}_{\perp}\cdot\bm{\sigma}),\label{MIdef}
\end{equation}
parameterized by a pair of three-dimensional unit vectors $\bm{\nu}$ and $\bm{n}_{\perp}$. The vector $\bm{n}_{\perp}$ should be orthogonal to $\bm{n}$ but $\bm{\nu}$ is not so constrained. Three common types of confinement are the zigzag edge, with $\bm{\nu}=\pm\bm{\hat{z}}$, $\bm{n}_{\perp}=\bm{\hat{z}}$; the armchair edge, with $\bm{\nu}\cdot\bm{\hat{z}}=0$, $\bm{n}_{\perp}\cdot\bm{\hat{z}}=0$; and infinite mass confinement, with $\bm{\nu}=\bm{\hat{z}}$, $\bm{n}_{\perp}\cdot\bm{\hat{z}}=0$.

To determine the edge states we consider a local coordinate system such that the boundary is along the $y$-axis (so $\bm{n}=-\bm{\hat x}$), and we choose a local gauge such that $\bm{A}=Bx\bm{\hat y}$. The wave number $q$ along the boundary is then a good quantum number. In order to simplify the notation we measure energies in units of $\hbar v/l_{m}$ and lengths in units of $l_{m}$. (Units will be reinstated in the final results.) Eigenstates of Eq.\ (\ref{DBdG2}) that decay for $x\rightarrow\infty$ have the form
\begin{eqnarray}
&&\Psi(x,y)=e^{iqy}\begin{pmatrix}
C_{e}\otimes\Phi_{e}(x+q)\\
C_{h}\otimes\Phi_{h}(x-q)
\end{pmatrix},\label{eigenstate}\\
&&\Phi_{e}(\xi)=e^{-\tfrac{1}{2}\xi^{2}}\begin{pmatrix}
-i(\mu+\varepsilon)H_{(\mu+\varepsilon)^{2}/2-1}(\xi )\\
H_{(\mu+\varepsilon)^{2}/2}(\xi )
\end{pmatrix},\label{Phie}\\
&&\Phi_{h}(\xi )=e^{-\tfrac{1}{2}\xi ^{2}}\begin{pmatrix}
H_{(\mu-\varepsilon)^{2}/2}(\xi )\\
-i(\mu-\varepsilon)H_{(\mu-\varepsilon)^{2}/2-1}(\xi )
\end{pmatrix},\label{Phih}
\end{eqnarray}
in the region $x>0$ (where $\Delta\equiv 0$). The function $H_{\alpha}(x)$ is the Hermite function. The two-component spinors $C_{e}$ and $C_{h}$ determine the valley isospin of the electron and hole components, respectively.

\begin{figure}[tb]
\centerline{\includegraphics[width=0.9\linewidth]{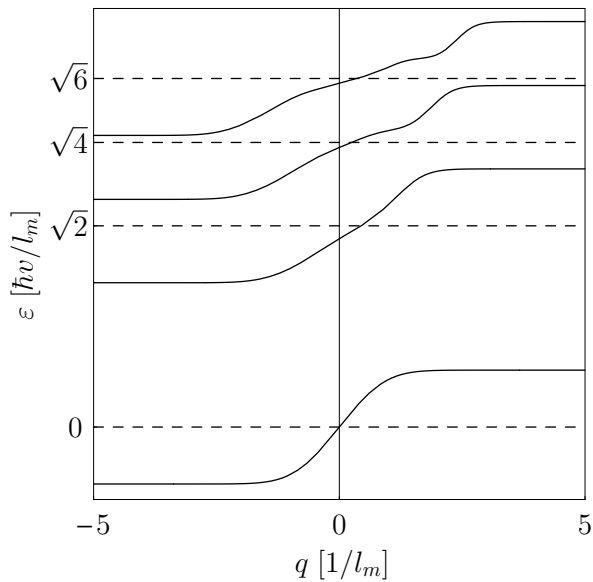}}
\caption{\label{fig_NS_dispersion}
Dispersion relation of edge states in graphene along the normal-superconducting interface, calculated from Eq.\ (\ref{NSdispersion}) for $|\varepsilon|\ll\Delta_{0}$. The dotted lines are for $\mu=0$, the solid lines for $\mu=0.4\,\hbar v/l_{m}$.
}
\end{figure}

The dispersion relation between energy $\varepsilon$ and momentum $q$ follows by substitution of the state (\ref{eigenstate}) into the boundary condition (\ref{calMdef}). At the NS interface we take Eq.\ (\ref{MNSdef}) for the boundary condition and obtain
\begin{eqnarray}
&&f_{\mu+\varepsilon}(q)-f_{\mu-\varepsilon}(-q)=\frac{\varepsilon[f_{\mu+\varepsilon}(q)f_{\mu-\varepsilon}(-q)+1]}{\sqrt{\Delta_{0}^{2}-\varepsilon^{2}}},\nonumber\\
&&f_{\alpha}(q)=\frac{H_{\alpha^2/2}(q)}{\alpha H_{\alpha^2/2-1}(q)}.\label{NSdispersion}
\end{eqnarray}
The solutions $\varepsilon_{n}(q)$, numbered by a mode index $n=0,\pm 1,\pm 2,\ldots$, are plotted in Fig.\ \ref{fig_NS_dispersion}. Notice that the dispersion relation has the inversion symmetry $\varepsilon(q)=-\varepsilon(-q)$. Each mode has a twofold valley degeneracy, because the boundary condition (\ref{MNSdef}) is isotropic in the valley isospin $\bm{\nu}$. The two degenerate eigenstates (labeled $\pm$) have $C_{e}^{\pm}=c_{e}|\pm\bm{\nu}\rangle$, $C_{h}^{\pm}=c_{h}|\pm\bm{\nu}\rangle$, with $|\pm\bm{\nu}\rangle$ eigenstates of $\bm{\nu}\cdot\bm{\tau}$ \cite{note3}.

\begin{figure}[tb]
\centerline{\includegraphics[width=0.9\linewidth]{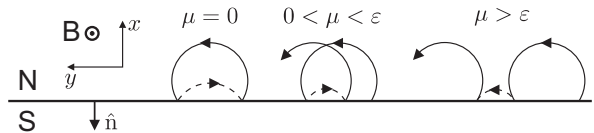}}
\caption{\label{fig_trajectories}
Cyclotron orbits of Andreev reflected electrons and holes.
}
\end{figure}

The expectation value $v_{n}=\hbar^{-1}d\varepsilon_{n}/dq$ of the velocity along the boundary in the $n$-th mode is determined by the derivative of the dispersion relation. We see from Fig.\ \ref{fig_NS_dispersion} that the edge states all propagate in the same direction, dictated by the sign of $B$ and $\mu$. The velocity vanishes for $|q|\rightarrow\infty$, as the NS edge states evolve into the usual dispersionless Landau levels deep in the normal region. For $q\rightarrow-\infty$ the Landau levels contain electron excitations at energy $\varepsilon_{n}=\sqrt{2}(\hbar v/l_{m})\,{\rm sign}\,(n)\sqrt{|n|}-\mu$, while for $q\rightarrow\infty$ they contain hole excitations with $\varepsilon_{n}=\sqrt{2}(\hbar v/l_{m})\,{\rm sign}\,(n)\sqrt{|n|}+\mu$. For $\mu=0$ the NS edge states have zero velocity at any $q$ for $|\varepsilon|\ll\Delta_{0}$. As illustrated in Fig.\ \ref{fig_trajectories}, the localization of the edge states as $\mu\rightarrow 0$ happens because for $|\varepsilon|>|\mu|$ the electron and hole excitations move in opposite directions along the boundary, while for $|\varepsilon|<|\mu|$ they move in the same direction.

\begin{figure}[tb]
\centerline{\includegraphics[width=0.9\linewidth]{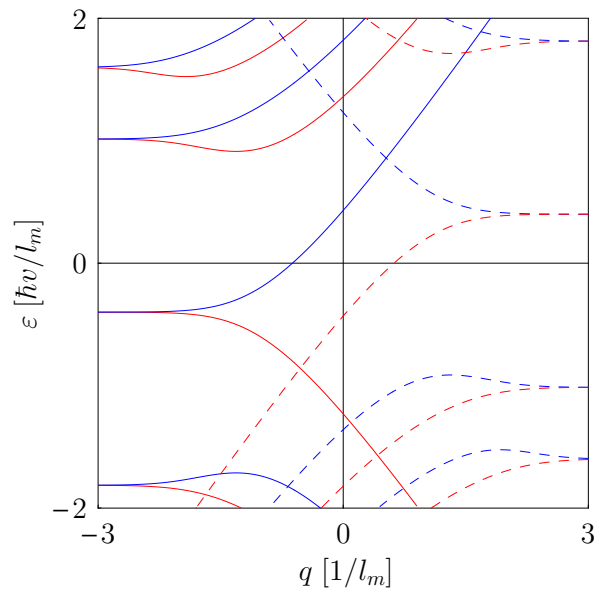}}
\caption{\label{I_Dispersion}
Dispersion relation of states along the insulating edge, calculated from Eqs.\ (\ref{epsplus}) and (\ref{epsminus}) for $\mu=0.4\,\hbar v/l_{m}$ and $\theta=\pi/2$. The solid lines are the electron states (blue $\varepsilon_{e}^{+}$, red $\varepsilon_{e}^{-}$), the dashed lines are the hole states (blue $\varepsilon_{h}^{+}$, red $\varepsilon_{h}^{-}$).
}
\end{figure}

Turning now to the insulating edge, we take the boundary condition (\ref{MIdef}). For an edge along the $y$-axis we have $\bm{n}_{\perp}=(0,\sin\theta,\cos\theta)$. The valley degeneracy is broken in general, with different dispersion relations for the two eigenstates $|\pm\bm{\nu}\rangle$ of $\bm{\nu}\cdot\bm{\tau}$. The dispersion relations for electrons  and holes are related by $\varepsilon^{\pm}_{h}(q)=-\varepsilon^{\mp}_{e}(-q)$. For sufficiently small $\mu$ there is one electron and one hole state at the Fermi level, of opposite isospins. (Note that electrons and holes from the {\em same\/} valley have {\em opposite\/} isospins.) We fix the sign of $\bm{\nu}$ such that $|+\bm{\nu}\rangle$ is the electron eigenstate and $|-\bm{\nu}\rangle$ the hole eigenstate. We find that $\varepsilon^{+}_{e}(q)$ is determined by the equation
\begin{equation}
f_{\mu+\varepsilon}(q)=\tan(\theta/2),\label{epsplus}
\end{equation}
while $\varepsilon^{-}_{e}(q)$ is determined by
\begin{equation}
f_{\mu+\varepsilon}(q)=-{\rm cotan}\,(\theta/2).\label{epsminus}
\end{equation}
The dispersion relations plotted in Fig.\ \ref{I_Dispersion} are for the case $\theta=\pi/2$ of an armchair edge. The case $\theta=0$ of a zigzag edge contains additional dispersionless states away from the Fermi level \cite{Bre06}, but these play no role in the electrical conduction.

To determine the conductance $G_{\rm NS}$ we need to calculate the transmission matrix $t$ of the edge states at the Fermi level. Edge states approach the superconductor along the insulating edge ${\rm I}_{1}$ (with parameters $\bm{\nu}_{1},\theta_{1}$), then propagate along the NS interface, and finally return along the insulating edge ${\rm I}_{2}$ (with parameters $\bm{\nu}_{2},\theta_{2}$). At sufficiently small $\mu$ each insulating edge ${\rm I}_{p}$ supports only two propagating modes, one electron mode $\propto|+\bm{\nu}_{p}\rangle$ and one hole mode $\propto|-\bm{\nu}_{p}\rangle$. The NS interface also supports two propagating modes at small $\mu$, of mixed electron-hole character and valley degenerate. The conductance is given by \cite{Blo82}
\begin{equation}
G_{\rm NS}=\frac{2e^{2}}{h}(1-T_{ee}+T_{he})=\frac{4e^{2}}{h}T_{he},\label{GNSdef}
\end{equation}
with $T_{ee}=|t_{++}|^{2}$ the probability that an electron incident along ${\rm I}_{1}$ returns along ${\rm I}_{2}$ as an electron and $T_{he}=|t_{-+}|^{2}$ the probability that the electron returns as a hole. Because electrons and holes cannot enter into the superconductor, these two probabilities must add up to unity --- hence the second equality in Eq.\ (\ref{GNSdef}). (The factor of two accounts for the spin degeneracy.)

Since the unidirectional motion of the edge states prevents reflections, the transmission matrix $t$ from ${\rm I}_{1}$ to ${\rm I}_{2}$ is the product of the transmission matrices $t_{1}$ from ${\rm I}_{1}$ to NS and $t_{2}$ from NS to ${\rm I}_{2}$. Each of the matrices $t_{p}$ is a $2\times 2$ unitary matrix, diagonal in the basis $|\pm\bm{\nu}_{p}\rangle$:
\begin{equation}
t_{p}=e^{i\phi_{p}}|+\bm{\nu}_{p}\rangle\langle+\bm{\nu}_{p}|+e^{i\phi'_{p}}|-\bm{\nu}_{p}\rangle\langle-\bm{\nu}_{p}|.\label{tpdef}
\end{equation}
The phase shifts $\phi_{p},\phi'_{p}$ need not be determined. Using $|\langle\bm{\nu}_{1}|\pm\bm{\nu}_{2}\rangle|^{2}=\frac{1}{2}(1\pm\bm{\nu}_{1}\cdot\bm{\nu}_{2})$, we obtain from $t=t_{2}t_{1}$ the required transmission probabilities
\begin{equation}
T_{he}=1-T_{ee}=\tfrac{1}{2}(1-\bm{\nu}_{1}\cdot\bm{\nu}_{2}).\label{tpmresult}
\end{equation}
Substitution into Eq.\ (\ref{GNSdef}) gives our central result (\ref{GNSresult}).

Referring to Fig.\ \ref{NS_layout}, we see that $G_{\rm NS}=0$ in the case (a) of a superconducting contact to a single edge ($\bm{\nu}_{1}=\bm{\nu}_{2}$) --- regardless of whether the edge is zigzag or armchair. In the case (c) of a contact between a zigzag and an armchair edge we have $\bm{\nu}_{1}\cdot\bm{\nu}_{2}=0\Rightarrow G_{\rm NS}=2e^{2}/h$. The case (b) of a contact between two opposite edges has $\bm{\nu}_{1}=-\bm{\nu}_{2}\Rightarrow G_{\rm NS}=4e^{2}/h$ if both edges are zigzag; the same holds if both edges are armchair separated by a multiple of three hexagons (as in the figure); if the number of hexagons separating the two armchair edges is not a multiple of three, then $\bm{\nu}_{1}\cdot\bm{\nu}_{2}=1/2\Rightarrow G_{\rm NS}=e^{2}/h$.

Intervalley relaxation at a rate $\Gamma$ tends to equalize the populations of the two degenerate modes propagating along the NS interface. This becomes appreciable if $\Gamma L/v_{0}\gtrsim 1$, with $L$ the length of the NS interface and $v_{0}=\hbar^{-1}d\varepsilon_{0}/dq\simeq\min(v/2,\sqrt{2}\,\mu l_{m}/\hbar)$ the velocity along the interface. The density matrix $\rho=\rho_{0}(1-e^{-\Gamma L/v_{0}})+\rho_{1}e^{-\Gamma L/v_{0}}$ then contains a valley isotropic part $\rho_{0}\propto\tau_{0}$ with $T_{ee}=T_{eh}=1/2$ and a nonequilibrium part $\rho_{0}\propto|\bm{\nu}_{1}\rangle\langle\bm{\nu}_{1}|$ with $T_{ee},T_{eh}$ given by Eq.\ (\ref{tpmresult}). The conductance then takes the form 
\begin{equation}
G_{\rm NS}=\frac{2e^{2}}{h}\bigl(1-e^{-\Gamma L/v_{0}}\cos\Theta\bigr).\label{tpmrelaxation}
\end{equation}
A nonzero conductance when the supercurrent covers a single edge ($\Theta=0$) is thus a direct measure of the intervalley relaxation.

In conclusion, we have shown that the valley structure of quantum Hall edge states in graphene, which remains hidden in the Hall conductance, can be extracted from the current that flows through a superconducting contact. Since such contacts have now been fabricated succesfully \cite{Hee06,Sha06}, we expect that this method to detect valley polarization can be tested in the near future.  

This research was supported by the Dutch Science Foundation NWO/FOM.


\begin{thebibliography}{99}
\bibitem{note1} The operators (\ref{Hdef}) and (\ref{Tdef}) in the valley isotropic basis are related to their counterparts in Ref.\ \onlinecite{Bee06} by the unitary transformation $H\rightarrow UHU^{\dagger}$, $T\rightarrow UTU^{\dagger}$, with $U=\tfrac{1}{2}(\tau_{0}+\tau_{z})\otimes\sigma_{0}+\tfrac{1}{2}(\tau_{0}-\tau_{z})\otimes\sigma_{x}$. Since $\Delta$ is a scalar, it remains unchanged by this transformation.
\bibitem{note2} Without the restriction to time-reversal symmetry the most general form of $M_{\rm I}$ is $M_{\rm I}=\tau_{0}\otimes\bigl((\bm{n}_{\perp}\times\bm{n})\cdot\bm{\sigma}\bigr)\cos\alpha+(\bm{\nu}\cdot\bm{\tau})\otimes(\bm{n}_{\perp}\cdot\bm{\sigma})\sin\alpha$. This four-parameter family of boundary conditions is more general than the three-parameter family of Ref.\ \cite{McC04}.
\bibitem{note3} The coefficients $c_{e,h}$ are given by $c_{e}/c_{h}=(\mu-\varepsilon) H_{(\mu-\varepsilon)^2/2-1}(-q)/(iH_{(\mu+\varepsilon)^2/2}(q)\cos\beta +i(\mu+\varepsilon) H_{(\varepsilon +\mu )^2/2-1}(q) \sin\beta$).
\end{thebibliography}
\end{document}